\documentclass[12pt,twocolumn,twoside,lineno]{pnas-new}
\usepackage{graphicx}
\templatetype{pnasmathematics} 

\title{Locating Order-Disorder Phase Transition in a Cardiac System}

\author[*a]{Hiroshi Ashikaga}
\author[b]{Ameneh Asgari-Targhi} 

\affil[a]{Cardiac Arrhythmia Service, Johns Hopkins University School of Medicine, 600 N Wolfe Street, Carnegie 568, Baltimore, Maryland 21287, USA}
\affil[b]{Division of Sleep and Circadian Disorders, Brigham and Women’s Hospital, 221 Longwood Avenue, Suite 438, Boston, Massachusetts 02115, USA}

\leadauthor{Ashikaga} 

\authordeclaration{Neither author declares any conflict of interest.}
\correspondingauthor{\textsuperscript{*}To whom correspondence should be addressed. E-mail: hashika1@jhmi.edu}

\keywords{Complex systems $|$ Phase transition $|$ Information theory $|$ Cardiac dynamics} 

\begin{abstract} 
To date, scientific investigation of order-disorder phase transitions has focused on discovering early warning signs to predict when the transition occurs. However, little attention has been paid to where in the networked dynamical system the phase transition begins. The imminent phase transition may be mitigated or prevented by identifying and modifying the specific components of the system responsible for the initiation of the transition. Here we present an information-theoretic approach to predicting the locations of an order-disorder phase transition from a regular heart rhythm to fibrillation in a cardiac system. We demonstrate the effectiveness of our approach by applying it to numerical simulations of a two-dimensional cardiac system. We show that, by analyzing communication between the components of the system, information-theoretic metrics such as channel capacity, mutual information, and transfer entropy can predict geometrical borders beyond which an order-disorder phase transition from a regular heart rhythm to fibrillation occurs. Importantly, we find that channel capacity and mutual information progressively decline and reach zero at the border of phase transition. This indicates that those information-theoretic metrics can serve as order parameters to describe the macroscopic behavior of the system. Our approach is computationally efficient and is applicable to many complex systems of interest in distinct physical, chemical, and biological disciplines. Our approach could ultimately contribute to an improved therapy of clinical conditions such as sudden cardiac death by identifying potential targets of interventional therapies.
\end{abstract}

\dates{This manuscript was compiled on \today}

\begin{document}

\verticaladjustment{-2pt}

\maketitle
\thispagestyle{firststyle}
\ifthenelse{\boolean{shortarticle}}{\ifthenelse{\boolean{singlecolumn}}{\abscontentformatted}{\abscontent}}{}

\dropcap{P}hase transitions between ordered and disordered states are pervasive in complex systems consisting of large numbers of interacting components \citep{buhl2006disorder,miramontes1995order}. Examples of phase transitions include stock market crash \citep{may2008complex}, climate change \citep{lenton2008tipping}, ecological collapse \citep{scheffer2001catastrophic}, population dynamics \citep{dai2012generic,dai2013slower}, epileptic seizures \citep{kramer2012human}, and flocking of birds \citep{Christodoulidi2014flocking}. In biological systems, phase transitions are particularly common because they tend to self-organize towards criticality, the border between order and disorder \citep{mora2011biological,hesse2014self,hidalgo2014information}. Previous work on phase transitions primarily focused on discovering early warning signs to predict when the phase transition occurs \citep{scheffer2009early}. However, little attention has been paid to where in the networked dynamical system the order-disorder phase transition begins, which is at least as important as when it occurs. The imminent phase transition may be mitigated or prevented by identifying and modifying the specific components of the system responsible for the initiation of the phase transition. 

In the heart, fibrillation is a disordered state of cardiac excitation that is one of the most common causes of sudden cardiac death, accounting for an estimated 15-20$\%$ of all deaths worldwide \citep{hayashi2015spectrum}. Fibrillation is frequently initiated by a train of rapid stimuli from a region of the heart other than the sinoatrial node. Common clinical examples include electrical misfiring from the pulmonary vein inducing atrial fibrillation \citep{haissaguerre1998spontaneous}, or ectopic beats from the Purkinje fibers causing ventricular fibrillation \citep{haissaguerre2016ventricular}. Identifying the components that initiate an order-disorder phase transition from a regular heart rhythm to fibrillation enables an interventional strategy to targeting those culminating components to prevent morbidity and mortality resulting from atrial \citep{verma2015approaches} and ventricular fibrillation \citep{knecht2009long,cheniti2017vf,krummen2015modifying}.

Information theory can be used to locate the network component that determines the behavior of the system \citep{quax2013diminishing}. In addition, information-theoretic metrics such as mutual information \citep{1948:shannon01} and transfer entropy \citep{schreiber2000measuring} can be used to predict phase transitions in many complex systems \citep{matsuda1996mutual,gu2007universal,vicsek1995novel,wicks2007mutual,ribeiro2008mutual,harre2009phase, barnett2013information}. However, the applicability of these metrics in complex systems has two significant limitations. First, these metrics require measuring the time series history of all the components in the system, which is virtually intractable in real-world systems. Second, due to the large number of components, it is not trivial to calculate those information-theoretic measures between all-to-all pairs of components in the system. 

The aim of this work was to develop a computationally efficient, information-theoretic approach to predict the locations of an order-disorder phase transition from a regular heart rhythm to fibrillation in a cardiac system. Since fibrillation is initiated by a wavebreak, which is an intersection between the wavefront (action potential upstroke) and the waveback (action potential repolarization) of electrical traveling waves \citep{weiss2005dynamics}, we predicted the locations of phase transition based on our information-theoretic approach, and compared those locations with the measured locations of wavebreaks. 

To simulate the common clinical examples of initiation of fibrillation, we consider a cardiac system where the system behavior is controlled by one single driving component that fires a train of rapid stimuli. In this setting, mutual information quantifies shared information, or the joint probability distribution (macrostate) over the possible microstates, between the driving component and each component of the system. Transfer entropy quantifies information flow from the driving component to each component of the system. In essence, our approach obviates the need for accounting for interactions between all-to-all pairs of components in the system, but instead only requires a communication analysis between the driving component and each component of the system, which is computationally feasible in real-world cases \citep{ashikaga2015modelling}. 

Fibrillation is often preceded by alternating action potential duration (APD) called APD alternans \citep{guevara1984electrical,quail2015predicting}. APD alternans is an intrinsic oscillatory dynamics of cardiac cells that arises out of a period-doubling bifurcation in the stochastic dynamics of intracellular calcium cycling \citep{restrepo2008calsequestrin,restrepo2009spatiotemporal,rovetti2010spark}. The emergence of APD alternans could be considered as an order-disorder phase transition \citep{alvarez2015calcium}, and earlier small clinical studies showed it may be associated with sudden cardiac death \citep{rosenbaum1994electrical}. However, a recent larger multicenter clinical trial concluded that APD alternans does not predict sudden cardiac death \citep{gold2008role}. In this work, although we analyze APD alternans as a key precursor to a phase transition, we focus on the emergence of fibrillation as an order-disorder phase transition since it is more clinically relevant \citep{goldberger2016sublimation}.

\section*{Results}

\subsection*{Conceptual overview}

\begin{figure}[!h]
  \centering
  \includegraphics[width=\linewidth,trim=0cm 21cm 20cm 0cm,clip]{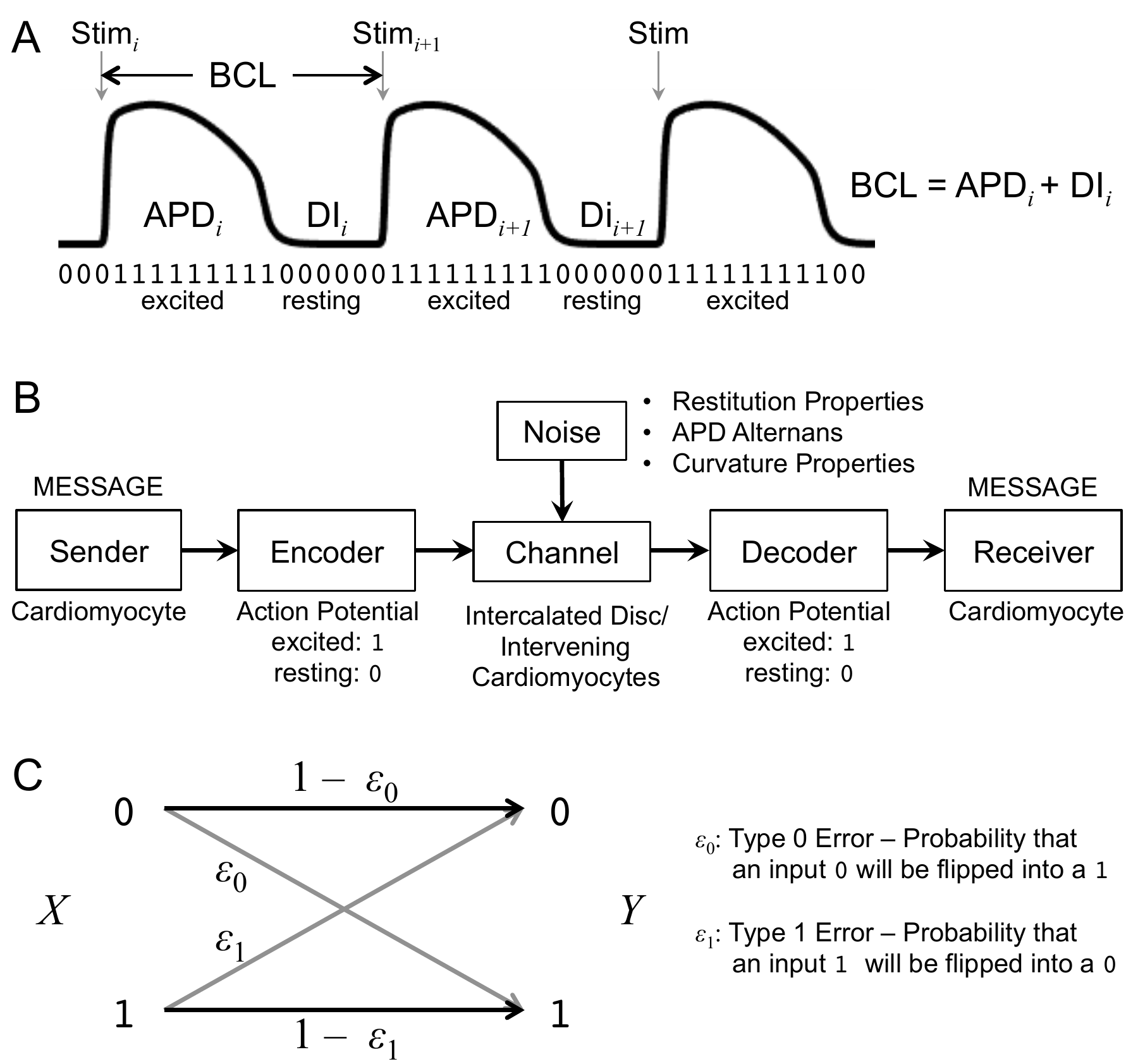}
  \caption{
    \textbf{Information-theoretic analysis of cardiac dynamics.}
    \textit{A. Cardiac action potential.} Three cardiac action potentials in response to regular stimuli with an external current ('\textit{Stim}') are shown. The basic cycle length (\textit{BCL}) is the interval between regular stimuli. The action potential duration (\textit{APD}) is measured at 90$\%$ repolarization ($APD_{90}$). The $i$th diastolic interval (\textit{DI$_i$}) is defined as the difference between BCL and the $i$th APD (\textit{APD$_i$}). The microstate of each cardiac component is encoded as 1 when excited (during APD) or 0 when resting (during DI). \textit{B. Heart as a communication system.} The cardiomyocytes act as an information source/encoder and a receiver/decoder with a channel being intercalated discs/intervening cardiomyocytes. Intrinsic dynamic cardiac properties such as restitution properties, APD alternans, and curvature properties serve as noise in the channel that could potentially introduce communication error. Figure modified from ~\citep{1948:shannon01}. \textit{C. Binary asymmetric channel.} The channel has a probability $\varepsilon_0$ that an input 0 will be flipped into a 1 (type 0 error) and a probability $\varepsilon_1$ for a flip from 1 to 0 (type 1 error).
  }
  \label{fig:conceptual}
\end{figure}

We performed an information-theoretic analysis of a cardiac system based on the conceptual method illustrated in Figure~\ref{fig:conceptual}. We describe the microstate of each cardiac component as either 1 (excited) or 0 (resting) in a binary time series (Figure~\ref{fig:conceptual}A). We use information theory to quantify the macrostate of each component and communication between components. From an information-theoretic perspective, intrinsic cardiac properties such as restitution properties, APD alternans, and curvature properties serve as noise in the communication channel that could potentially introduce communication error (Figure~\ref{fig:conceptual}B). We consider these channels to be a binary asymmetric channel, the most general form of binary discrete memoryless channel, to quantify the error and the channel capacity (Figure~\ref{fig:conceptual}C).

\subsection*{Restitution properties}

\begin{figure}[!h]
  \centering
  \includegraphics[width=\linewidth,trim=0cm 14cm 11cm 0cm,clip]{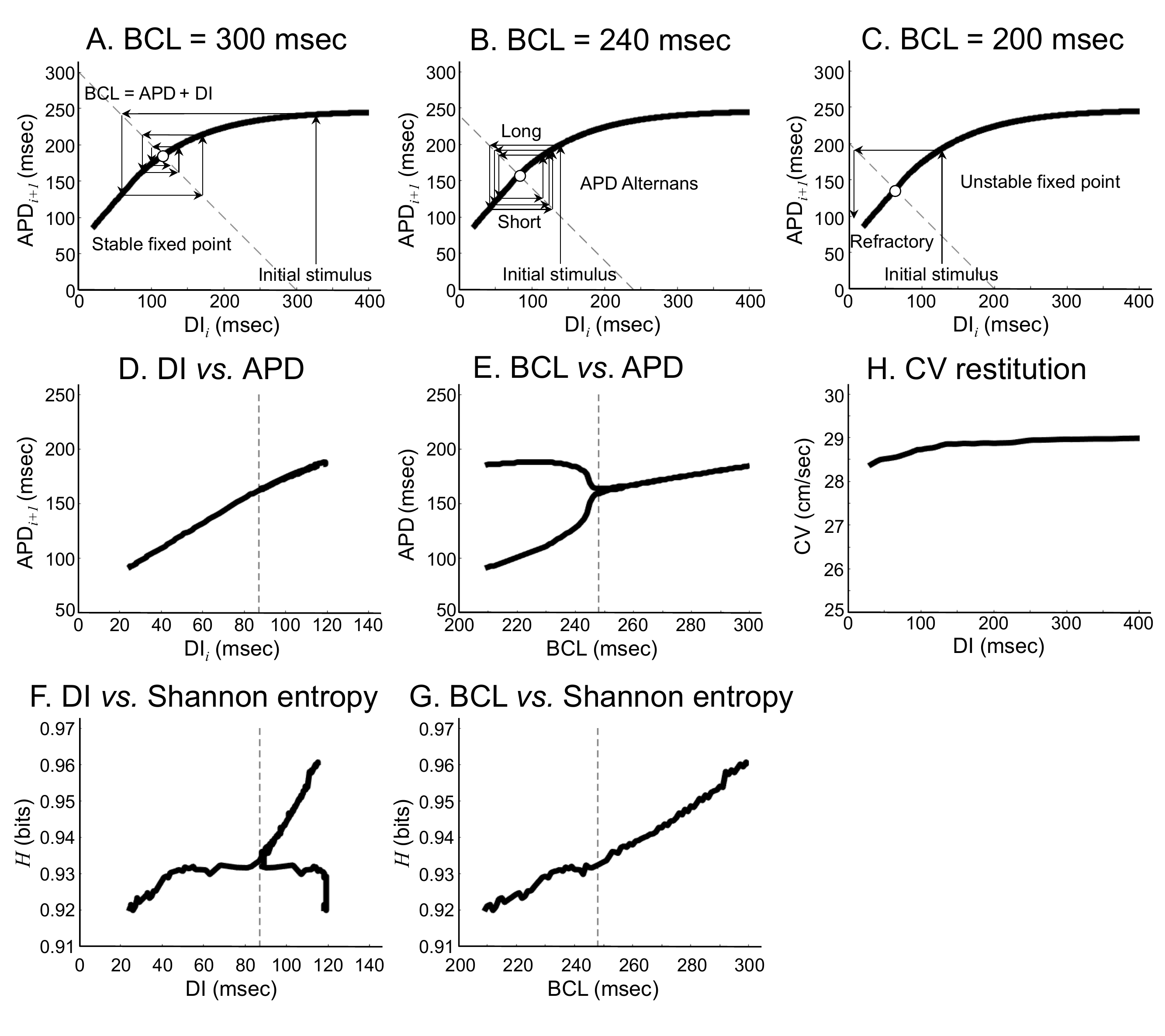}
  \caption{
    \textbf{Restitution properties.} \textit{A. BCL=300 msec (normal response).} The solid curve shows the APD restitution curve, and the dashed line shows the relationship between APD and DI for the new BCL (BCL=APD+DI). Over several stimuli, APD converges to a stable fixed point (APD=185 msec). \textit{B. BCL=240 msec (APD alternans).} The APD response shows oscillatory dynamics alternating with long (APD=184 msec) and short APD (APD=128 msec). \textit{C. BCL=200 msec (refractory response).} The APD response to the initial stimulus leads to DI shorter than the minimal value in the APD restitution curve. This results in a 2:1 response. Figure modified from ~\citep{2014zipes:aa}. \textit{D, E, F, and G.} A zoomed view of the steeper end of the APD restitution curve during BCL 200-300 msec. The dashed line shows the period-doubling bifurcation (BCL=248 msec). \textit{F} and \textit{G} show Shannon entropy as a function of DI and BCL, respectively. \textit{H. Conduction velocity (CV) restitution curve.}    
  }
  \label{fig:restitution}
\end{figure}

The APD restitution curve defines how the length of consecutive 0s (= diastolic interval, DI) determines the length of the next consecutive 1s (= APD) in the binary sequence of cardiac microstate. At BCL=300 msec, APD converges to a stable fixed point (APD=185 msec) after several beats (Figure~\ref{fig:restitution}A). At BCL=240 msec, the APD response shows a classic oscillatory dynamics of APD alternans with long (= 184 msec) and short APD (= 128 msec)(Figure~\ref{fig:restitution}B). At BCL=200 msec, every other stimulus cannot evoke APD because the component is refractory, resulting in a 2:1 response (Figure~\ref{fig:restitution}C). Overall, the model successfully reproduces the nonlinear, bifurcation behavior of cardiac system. As BCL progressively declines, the slope APD$/$DI increases and approaches 1 (Figure~\ref{fig:restitution}D, dashed line), when the cardiac microstate suddenly changes due to a period-doubling bifurcation. This corresponds to BCL=248 msec, beyond which APD alternans is observed (Figure~\ref{fig:restitution}E, dashed line). Shannon entropy shows a bifurcation when it is plotted against DI (Figure~\ref{fig:restitution}F, dashed line). This simply shows that one value of DI larger than the bifurcation represents two different cardiac macrostates that derive from two different BCL due to APD alternans. When Shannon entropy is plotted against BCL, it shows a small and steady decline across the bifurcation as BCL goes down (Figure~\ref{fig:restitution}G, dashed line). This indicates that the period-doubling bifurcation does not impact the cardiac macrostate, a probability distribution over the possible microstates. The model also successfully reproduces the conduction velocity (CV) restitution curve, which defines how the length of consecutive 0s (= DI) determines the delay of communication between two components (Figure~\ref{fig:restitution}D). CV is relatively flat for a wide range of DI, and only begins to decrease at a very short DI (approximately 100 msec).

\subsection*{APD alternans and conduction block in 1-D cable}

The model successfully reproduces important one-dimensional (1-D) properties of cardiac system, including APD alternans and conduction block. 

\begin{figure}[!h]
  \centering
  \includegraphics[width=\linewidth,trim=0cm 15cm 10cm 0cm,clip]{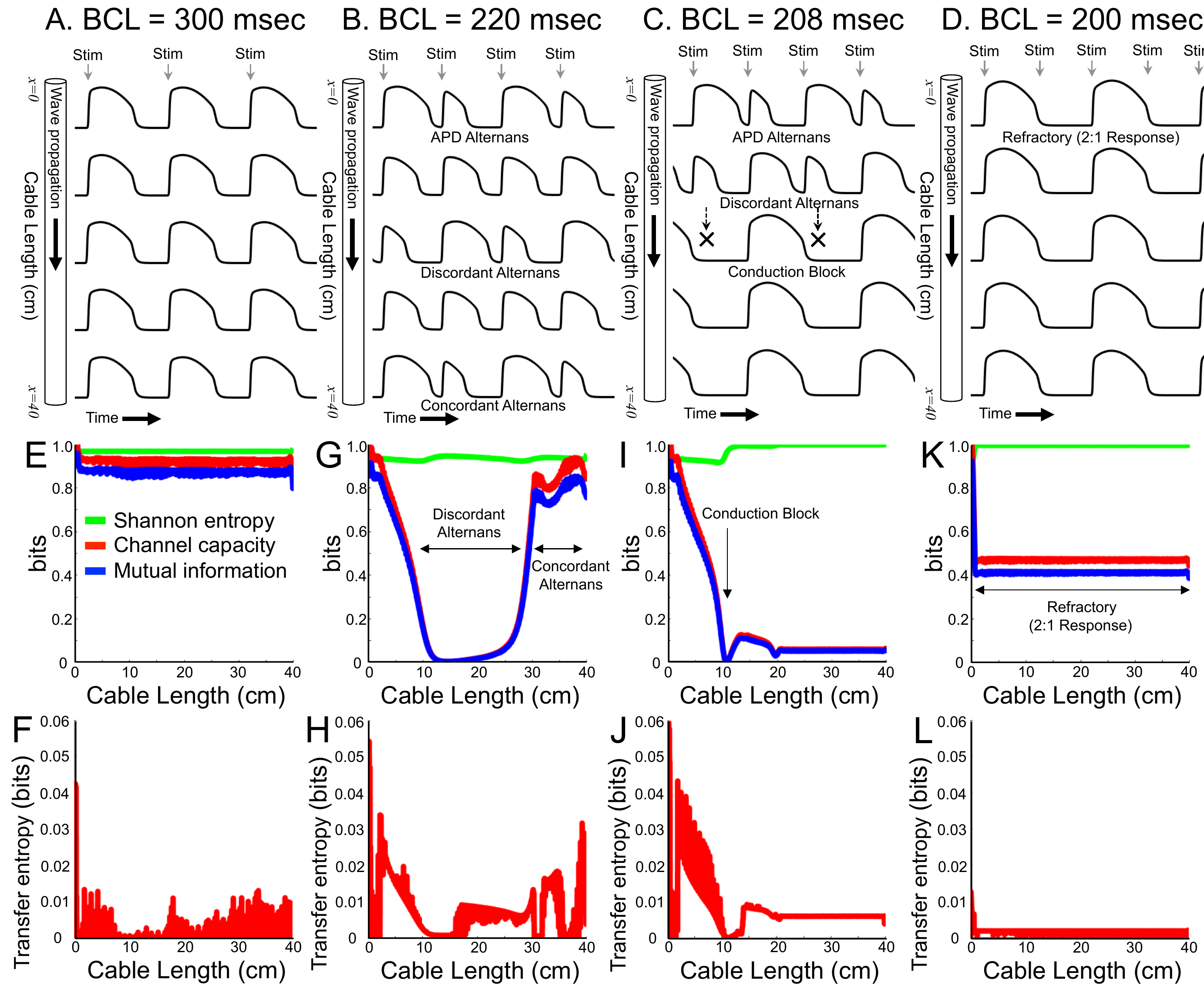}
  \caption{
    \textbf{APD alternans and conduction block in 1-D cable.} \textit{A. BCL=300 msec (normal response).} Waves of cardiac excitation travel from the stimulus site on the top ($x$=0 cm) to the bottom ($x$=40 cm) of a 40-cm, one-dimensional (1-D) cable illustrated on the left hand ($\Delta$x = 0.025 cm). \textit{B. BCL=220 msec (APD alternans).} \textit{C. BCL=208 msec (conduction block).} \textit{D. BCL=200 msec (refractory response).} Shannon entropy (bits, green line), channel capacity (bits, red line) and mutual information (bits, blue line) are shown as a function of cable length (cm) in \textit{E} (BCL=300 msec), \textit{G} (BCL=220 msec), \textit{I} (BCL=208 msec), and \textit{K} (BCL=200 msec). The stimulus site is at cable length = 0 cm. Transfer entropy (bits, red line) is shown as a function of cable length (cm) in \textit{F} (BCL=300 msec), \textit{H} (BCL=220 msec), \textit{J} (BCL=208 msec), and \textit{L} (BCL=200 msec). The stimulus site is at cable length = 0 cm. 
  }
  \label{fig:1d_cable}
\end{figure}

At BCL=300 msec, one stimulus generates one action potential, and APD is uniform spatially and temporally (APD=185 msec, Figure~\ref{fig:1d_cable}A). Shannon entropy is consistently high throughout the cable length (Figure~\ref{fig:1d_cable}E), which indicates that the cardiac macrostate is homogeneous across the cable. Channel capacity and mutual information are also consistently high. Mutual information is the lowest of the three information metrics but is close to channel capacity, which indicates that the channel is operating close to the maximum capacity. Transfer entropy  is small and variable with fluctuations across the cable length (Figure~\ref{fig:1d_cable}F).

At BCL=220 msec, one stimulus generates one action potential, but APD at the stimulus site is alternating with long (APD=188 msec) and short APD (APD=101 msec) (Figure~\ref{fig:1d_cable}B, top row). At a short distance away from the stimulus site on the cable (Figure~\ref{fig:1d_cable}B, second row), the APD alternans improves and APD becomes uniform temporally. This is called a node behavior since this site serves as a node to link the regions of concordant and discordant alternans. At midway of the cable (Figure~\ref{fig:1d_cable}B, third row), APD alternans reappears but $\pi$ out of phase from the stimulus site (\textit{discordant alternans}). At a further distance away from the stimulus site (Figure~\ref{fig:1d_cable}B, fourth row), the APD alternans improves and APD becomes uniform temporally again, showing another node behavior. At furthest distance away from the stimulus site (Figure~\ref{fig:1d_cable}B, bottom row), the APD alternans reappears in-phase with the stimulus site (\textit{concordant alternans}). Shannon entropy remains high and constant throughout the cable length (Figure~\ref{fig:1d_cable}G). Channel capacity is high near the stimulus site (cable length=0 cm), but steeply declines to near zero between cable length=10 cm and 25 cm, representing the region of discordant APD alternans. Beyond cable length=30 cm, channel capacity improves at $>$0.8 bits, reflecting concordant APD alternans. Mutual information follows the trend of the channel capacity, with only slightly lower values. Transfer entropy still fluctuates but declines steadily from the stimulus site and reaches the minimum value at the beginning of discordant alternans (cable length around 10 cm), indicating little information transfer from the source to this region (Figure~\ref{fig:1d_cable}H).
 
At BCL=208 msec, one stimulus generates one action potential with APD alternans at the stimulus site alternating with long (APD=185 msec) and short APDs (APD=90 msec) (Figure~\ref{fig:1d_cable}C, top row). At a short distance away from the stimulus site on the cable (Figure~\ref{fig:1d_cable}C, second row), the APD alternans becomes discordant, and eventually conduction block occurs (Figure~\ref{fig:1d_cable}C, third row). At midway and further, only one APD out of two stimuli is conducted (Figure~\ref{fig:1d_cable}C, fourth and bottom rows). There is a tendency for Shannon entropy to decline toward the site of conduction block, where it goes up again and remains high throughout the cable length (Figure~\ref{fig:1d_cable}I). Channel capacity is high near the stimulus site (cable length=0 cm), but steeply decreases to near zero at around cable length=10 cm, representing the region of discordant APD alternans and conduction block. Beyond the site of block, the channel capacity improves only slightly but continues to be low ($<$0.16 bits) throughout the cable. Mutual information follows the trend of the channel capacity, with only slightly lower values. Transfer entropy declines steadily but with fluctuations from the stimulus site and reaches the minimum value at the beginning of discordant alternans and conduction block (cable length around 10 cm)(Figure~\ref{fig:1d_cable}J). Beyond the site of conduction block, transfer entropy increases slightly and remains constant throughout the remaining cable length, indicating that the information of stimuli from the source continues to be transmitted to this region despite a 2:1 conduction.
 
At BCL=200 msec, BCL is shorter than APD, thus hits the refractory period of the previous APD (Figure~\ref{fig:1d_cable}D). As a result, only every other stimulus generates one action potential at the stimulus site (2:1 response). Shannon entropy remains high throughout the cable length, indicating that the 2:1 response does not impact the cardiac macrostate (Figure~\ref{fig:1d_cable}K). Channel capacity drops to approximately half of that of the origin (around 0.5 bits) and maintains the same level throughout the cable, reflecting the 2:1 response. Mutual information follows the trend of the channel capacity, with only slightly lower values. Transfer entropy remains close to zero throughout the cable length (Figure~\ref{fig:1d_cable}L), indicating little information transfer from the source when a 2:1 response is present.

\subsection*{Information dynamics in 1-D cable}

\begin{figure}[!h]
  \centering
  \includegraphics[width=\linewidth,trim=0cm 23cm 11cm 0cm,clip]{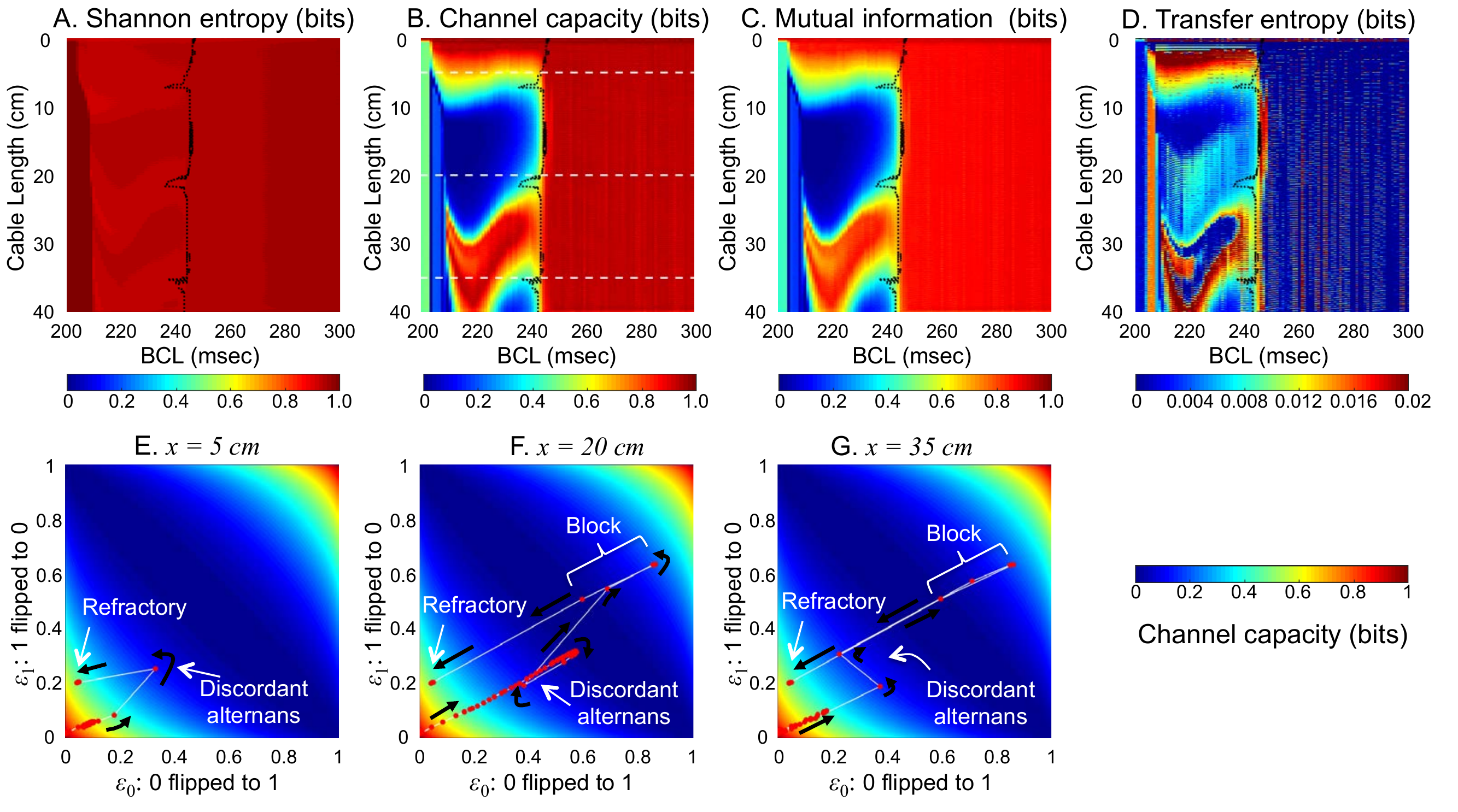}
  \caption{
    \textbf{Information dynamics in 1-D cable.} \textit{A. Shannon entropy.} Black dashed lines indicate a period-doubling bifurcation at each component with three local minima at $x$=7.1 cm (BCL=236 msec), $x$=21.5 cm (BCL=236 msec), and $x$=35.3 cm (BCL=237 msec). \textit{B. Channel capacity.} \textit{C. Mutual information.} \textit{D. Transfer entropy.} White dashed lines in \textit{B} indicate three components shown in the error space analysis, $x$=5cm \textit{E}, $x$=20cm \textit{F}, and $x$=35cm \textit{G}. In an error space, the $x$-axis is the probability $\varepsilon_0$ that an input 0 will be flipped into a 1 (type 0 error), and the $y$-axis is the probability $\varepsilon_1$ for a flip from 1 to 0 (type 1 error). Red dots, white lines and black arrows in the error space indicate the trajectory of each component as BCL goes down from 300 msec to 200 msec.
  }
  \label{fig:dynamics1d}
\end{figure}

Figure~\ref{fig:dynamics1d}A-D shows the dynamics of Shannon entropy, channel capacity, mutual information, and transfer entropy as a function of BCL ($x$-axis) and cable length ($y$-axis). Shannon entropy is robust against change in BCL and the location (Figure~\ref{fig:dynamics1d}A). This indicates that BCL has little impact on the cardiac macrostate. We also find that a period-doubling bifurcation is variable depending on the location of the cable (black dashed line). There are three local minima of bifurcation at $x$=7.1 cm (BCL=236 msec), $x$=21.5 cm (BCL=236 msec), and $x$=35.3 cm (BCL=237 msec). Channel capacity shows that those minima of bifurcation is located at the node between concordant and discordant APD alternans, indicating that the variability of bifurcation derives from the node behavior (Figure~\ref{fig:dynamics1d}B). To better understand the dynamics of cardiac macrostate as a function of BCL, we conducted an error-space analysis of each component where the $x$-axis is the probability $\varepsilon_0$ that an input 0 will be flipped into a 1 (type 0 error), and the $y$-axis is the probability $\varepsilon_1$ for a flip from 1 to 0 (type 1 error). As BCL declines from 300 msec to 200 msec, the cardiac component at $x$=5 cm starts at the origin of the error space and travels counterclockwise (Figure~\ref{fig:dynamics1d}E). Both $\varepsilon_0$ and $\varepsilon_1$ reaches a peak where discordant alternans is maximum before BCL hits the refractory period and shows a refractory response. At $x$=20 cm discordant alternans persists over many BCL before a conduction block occurs, and eventually a refractory response ensues (Figure~\ref{fig:dynamics1d}F). In contrast, at $x$=35 cm the trajectory is initially limited to the vicinity of the origin since concordant alternans occurs more frequently than discordant alternans (Figure~\ref{fig:dynamics1d}G). As BCL progressively declines, eventually a conduction block and a refractory response occur.

\subsection*{Order-disorder phase transition in 2-D lattice}

We applied regular stimuli at the top left component of a 2-D lattice, and used BCL as a control parameter to study the order-disorder phase transition in the cardiac system. We designed the model such that it is simple yet incorporates all of the intrinsic properties of 2-D cardiac dynamics, namely restitution properties, APD alternans and wavefront curvature properties (Figure~\ref{fig:conceptual}B). 

\begin{figure}[!h]
  \centering
  \includegraphics[width=0.8\linewidth,trim=0cm 20cm 18cm 0cm,clip]{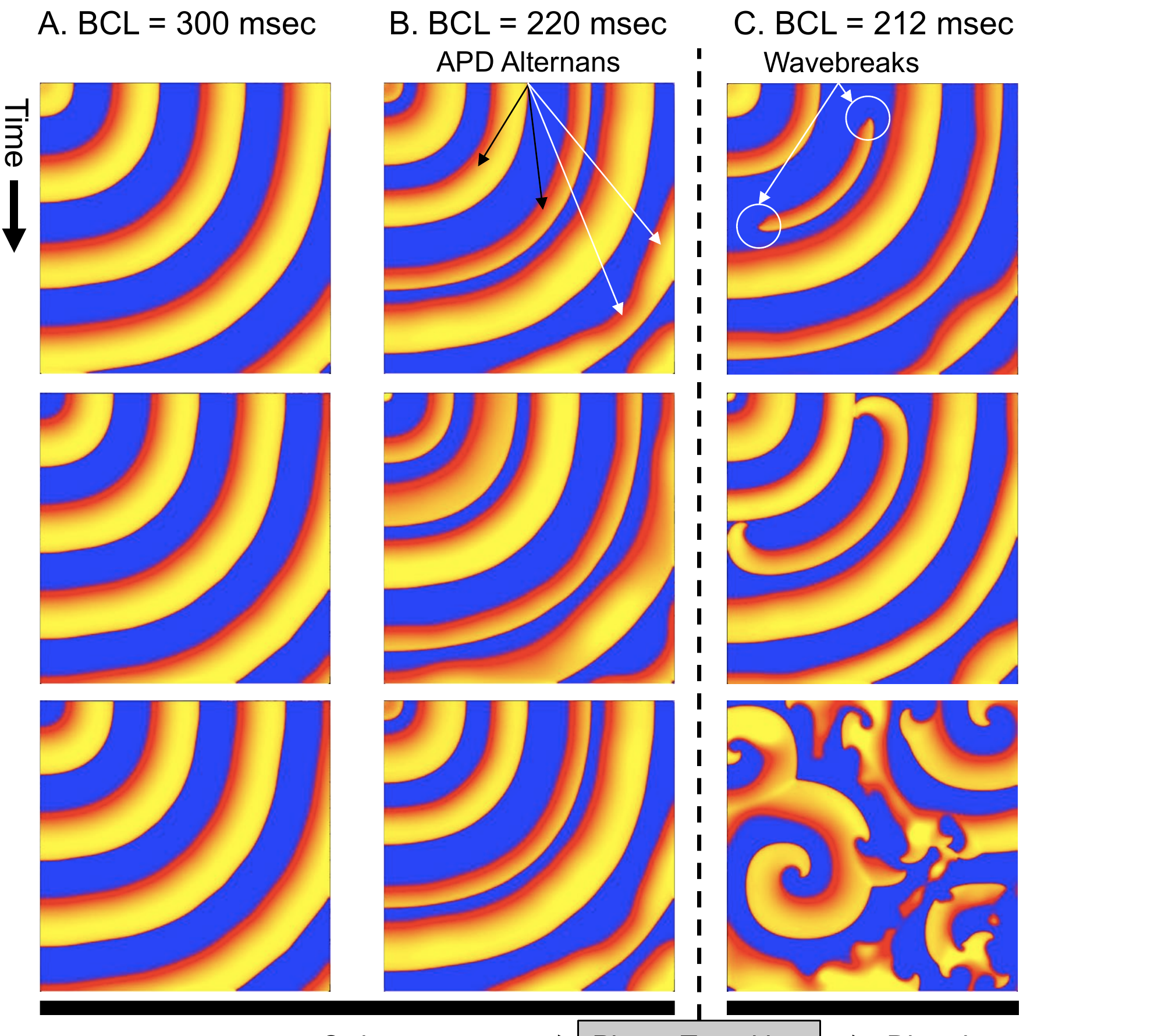}
  \caption{
    \textbf{Order-disorder phase transition in 2-D lattice.} The size of the lattice is 25cm $\times$ 25cm ($\Delta$x = 0.025 cm). The stimulus site is the top left component of the lattice, and the wave of cardiac excitation travels radially over the lattice. \textit{A. BCL=300 msec (normal APD response).} \textit{B. BCL=220 msec (APD alternans).} \textit{C. BCL=212 msec (wavebreaks).}
  }
  \label{fig:2d_lattice}
\end{figure}

At BCL=300 msec, each stimulus generates one action potential, and APD quickly becomes uniform spatially and temporally after a few stimuli (Figure~\ref{fig:2d_lattice}A). 

At BCL=220 msec, APD alternans is observed because BCL is shorter than the period-doubling bifurcation (Figure~\ref{fig:2d_lattice}B). Successive traveling waves alternate with long and short APD. APD alternans is also observed within the same wave as well as between waves. This is because the conduction velocity of the traveling wave is a function of the local wavefront curvature. The conduction velocity becomes slower as the local curvature becomes more convex because of a source-sink mismatch\citep{cabo1994wave}. This property makes the conduction velocity at the top and the left border of the lattice higher than that of the diagonal direction. In addition, this property makes the conduction velocity farther away from the stimulus site higher as the local curvature becomes less convex. Therefore, although the lattice is anatomically isotropic, it is functionally anisotropic depending on the wavefront curvature. 

At BCL=212 msec, partial conduction block occurs at the top and the left edges of the wave, creating wavebreaks (Figure~\ref{fig:2d_lattice}C). This is because the conduction velocity of these portions of the wave is higher than the rest of the wave. The wavebreaks initiate spiral waves that become completely out of sync with the stimuli at the top left corner of the lattice. Once initiated, the spiral waves break up into more spiral waves that persist until the end of the observation period. Such disordered states are observed at BCL equal to and shorter than 218 msec. We defined this sudden change in system behavior as an order-disorder phase transition in this study.

\subsection*{Information dynamics in 2-D lattice}

\begin{figure}[!h]
  \centering
  \includegraphics[width=0.7\linewidth,trim=0cm 10cm 18cm 0cm,clip]{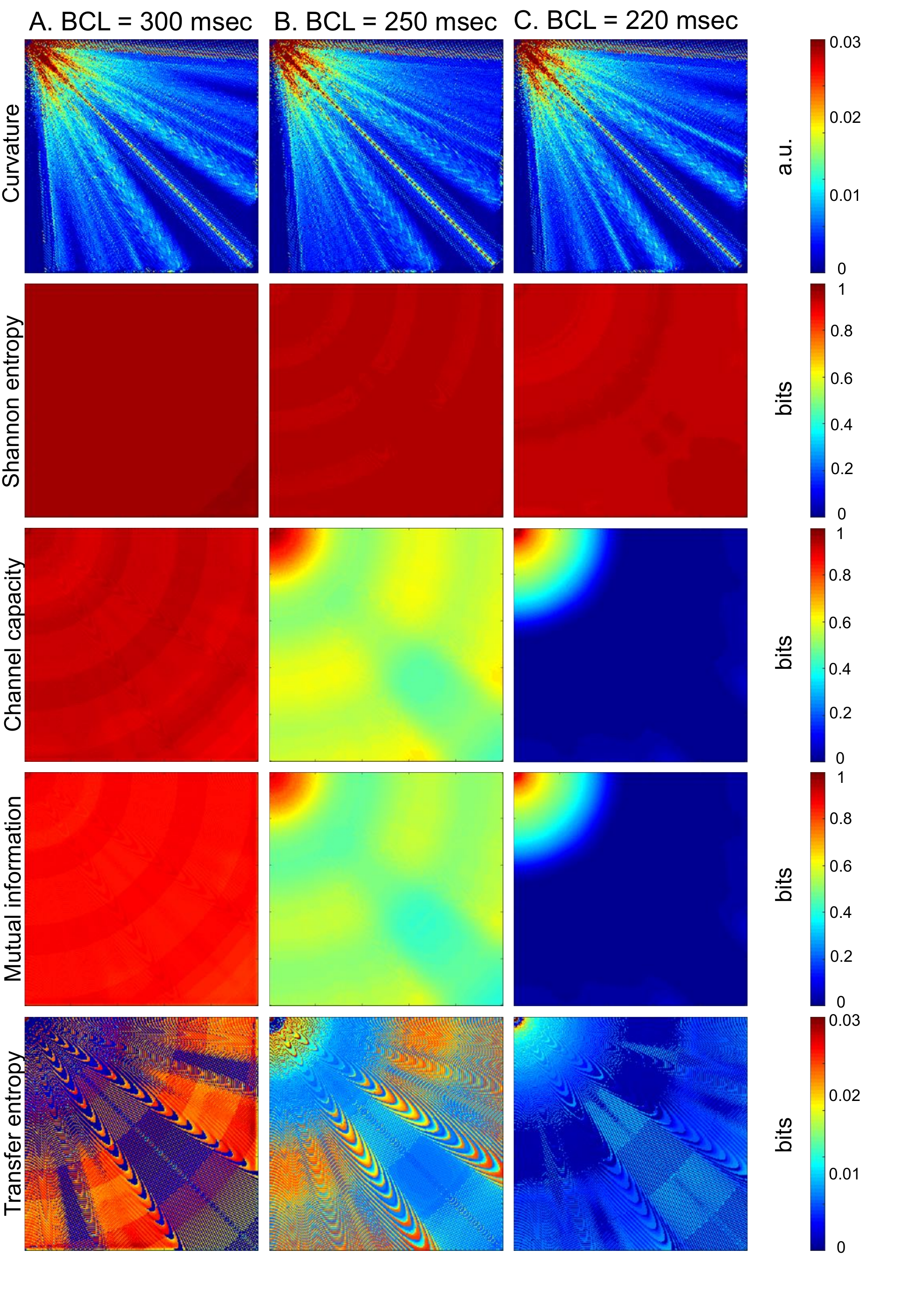}
  \caption{
    \textbf{Information dynamics in 2-D lattice.} Columns: \textit{A. BCL=300 msec (normal response).} \textit{B. BCL=250 msec (mild APD alternans).} \textit{C. BCL=220 msec (moderate APD alternans).} Rows: (from top to bottom) curvature (a.u.; arbitrary units), Shannon entropy (bits), channel capacity (bits), mutual information (bits), and transfer entropy (bits).
  }
  \label{fig:2d_profile}
\end{figure}

Wavefront curvature is regionally heterogeneous in the 2-D lattice (Figure~\ref{fig:2d_profile}, 'curvature'). Curvature was highest in the vicinity of the stimulus site, and declines farther away from the stimulus site as it becomes less convex. In addition, curvature is lowest at the top and the left border of the lattice because of the boundary effect. Importantly, curvature remains relatively invariant across different BCL. Shannon entropy also remains relatively invariant across different BCL, confirming little change in cardiac macrostate at different BCL (Figure~\ref{fig:2d_profile}, 'Shannon entropy'). As BCL decreases, channel capacity significantly declines because of discordant alternans except in the immediate vicinity of the stimulus site (Figure~\ref{fig:2d_profile}, 'Channel capacity'). Unlike the 1-D cable case discussed above, concordant alternans is not observed because the lattice is too small for concordant alternans to appear. At BCL=220 msec, channel capacity of the vast majority of the lattice is zero. Mutual information follows the trend of the channel capacity, with only slightly lower values (Figure~\ref{fig:2d_profile}, 'Mutual information'). Transfer entropy is more heterogeneous even at BCL=300 msec, and shows more complex dynamics than other information metrics (Figure~\ref{fig:2d_profile}, 'Transfer entropy'). In some regions, transfer entropy is low (dark blue color) at BCL=300 msec, which goes up at BCL=250 msec (light blue color), then goes down again at BCL=220 msec. This suggests that there is a complex and dynamic information transaction between the stimulus site and individual cardiac components.

\subsection*{Spatial heterogeneity of communication}

\begin{figure}[!h]
  \centering
  \includegraphics[width=0.7\linewidth,trim=0cm 9cm 18cm 0cm,clip]{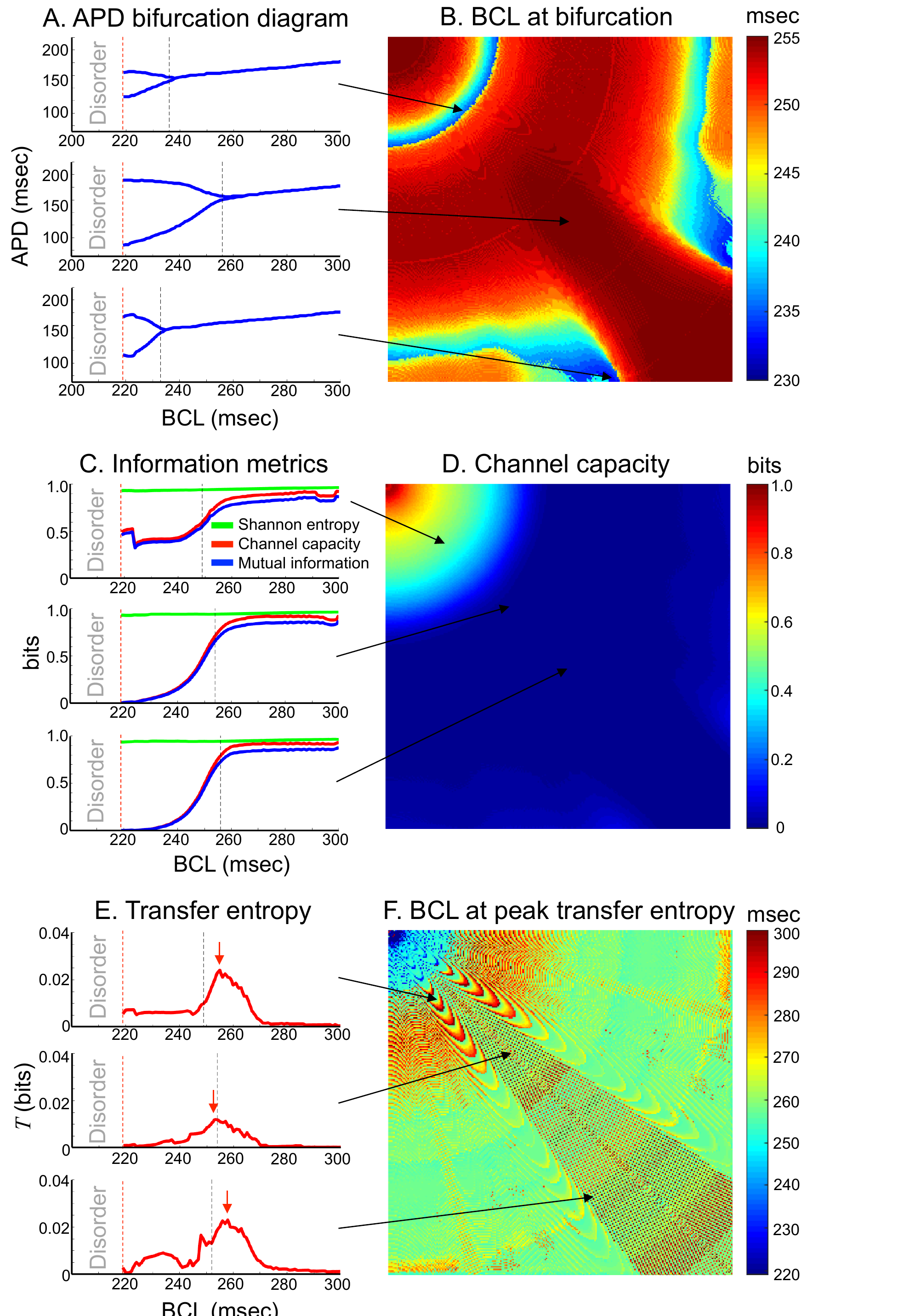}
  \caption{
    \textbf{Spatial heterogeneity of communication.} \textit{A. APD bifurcation diagram.} Dashed black lines indicate a period-doubling bifurcation at each component (BCL=230-255 msec). Dashed red lines indicate an order-disorder phase transition (BCL=218 msec). \textit{B. BCL (msec) at which period-doubling bifurcation occurs.} Note BCL at bifurcation is spatially heterogeneous although the system is anatomically homogeneous. Black arrows indicate the locations of components with the bifurcation diagrams on the left panels in \textit{A}. \textit{C. Information metrics between the stimulus site and each component.} Shannon entropy (green solid line, bits), channel capacity (red line, bits), and mutual information (blue line, bits). \textit{D. Channel capacity (bits) immediately prior to phase transition}. BCL=219 msec. Note the channel capacity is zero (= blue) in the vast majority of components in the 2-D lattice, except in the immediate vicinity of the stimulus site. Black arrows indicate the locations of components with the diagrams on the left panels in \textit{C}. \textit{E. Transfer entropy from the stimulus site to each component.} Red arrows indicate the peak of transfer entropy. \textit{F. BCL at peak transfer entropy.} Note the wide range of BCL that gives the peak transfer entropy. Black arrows indicate the locations of components with the diagrams on the left panels in \textit{E}. 
    }
  \label{fig:bifurcation}
\end{figure}

Similar to the result in the 1-D cable, BCL at which the period-doubling bifurcation occurs is spatially heterogeneous (Figure~\ref{fig:bifurcation}A, B). This result confirms that the system is functionally heterogeneous although anatomically homogeneous. The regions in transition between concordant, in-phase APD alternans and discordant, out-of-phase APD alternans show shorter BCLs at bifurcation because APD tends to be stable in these regions (node behavior). Importantly, all the components in the entire lattice reached BCL at bifurcation (BCL=230-255 msec) prior to the phase transition (BCL=218 msec). 

Shannon entropy remains relatively constant across the entire range of BCL at all the components in the 2-D lattice (Figure~\ref{fig:bifurcation}C). Similar to the result in the 1-D cable, the period-doubling bifurcation did not have any impact on Shannon entropy. Channel capacity remains relatively flat across the higher end of the BCL range, but it begins to decline as BCL declines and approaches the period-doubling bifurcation. Beyond the bifurcation the channel capacity progressively declines and reaches a plateau. The plateau is zero in the vast majority of components in the 2-D lattice, except in the immediate vicinity of the stimulus site (Figure~\ref{fig:bifurcation}D). The behavior of the mutual information was similar to that of the channel capacity, except that the channel capacity was only slightly higher than the mutual information.

Transfer entropy peaks at a wide range of BCL. Some components show that transfer entropy peaks at BCL longer than the period-doubling bifurcation (Figure~\ref{fig:bifurcation}E, top and bottom panels). In other components, transfer entropy peaks at BCL shorter than the period-doubling bifurcation (Figure~\ref{fig:bifurcation}E, middle panel). Overall, BCL at which transfer entropy peaks is spatially heterogeneous in the 2-D lattice (Figure~\ref{fig:bifurcation}F).

\subsection*{Locating order-disorder phase transition}

\begin{figure}[!h]
  \centering
  \includegraphics[width=0.8\linewidth,trim=0cm 9cm 15cm 0cm,clip]{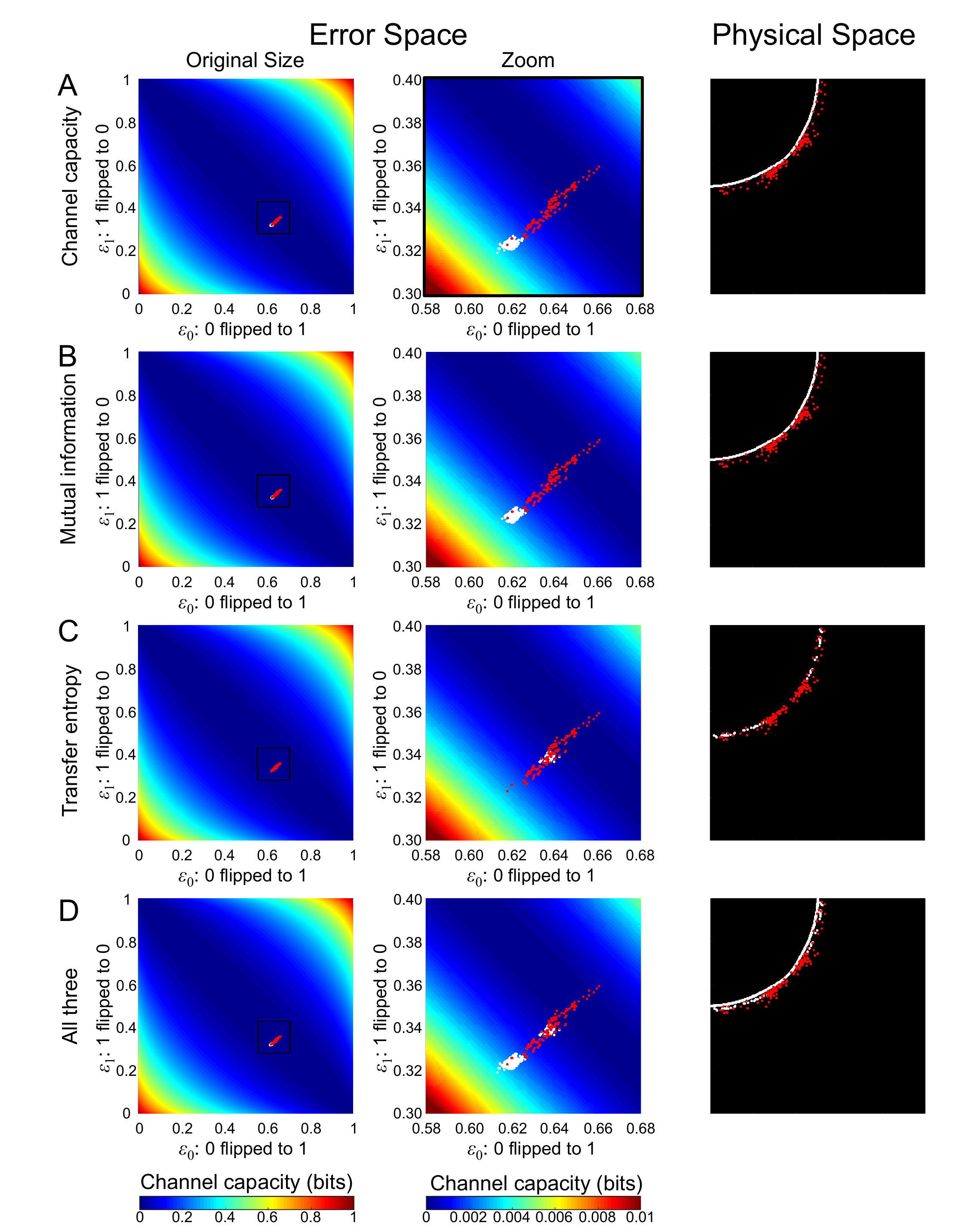}
  \caption{
    \textbf{Wavebreak localization} The figure shows predicted regions of an order-disorder phase transition (white) based on information-theoretic metrics immediately prior to the phase transition (BCL=219 msec) and actual wavebreak locations (red). The left (original size) and the middle columns (zoom) indicate the error space where the $x$-axis is the type 0 error (probability that an input 0 will be flipped into a 1), the $y$-axis is the type 1 error (probability that an input 1 will be flipped into a 0), and the color indicates the channel capacity at each location in the error space. The right column indicates the physical space of the 2-D lattice. \textit{A. Prediction based on channel capacity.} \textit{A. Prediction based on mutual information.} \textit{C. Prediction based on transfer entropy.} \textit{D. Prediction based on channel capacity, mutual information and transfer entropy.}
}
\label{fig:error2d}
\end{figure}

We predicted the region where an order-disorder phase transition begins according to a set of criteria of information metrics and compared the predicted region with wavebreak locations (Figure~\ref{fig:error2d}). We locate a total of 132 wavebreaks that initiate the first spiral wave in the system out of a total of 56 combinations of BCL train of stimuli. We use only the wavebreaks that initiate the first spiral wave because the first spiral wave induces a chain reaction of breaking up into multiple spiral waves \citep{fenton2002multiple}, and this reaction is an intrinsic cardiac dynamics that is independent of the interaction between the driving component and each component of the lattice.

Because channel capacity reaches zero at the order-disorder phase transition (Figure~\ref{fig:bifurcation}C), we used it as a criterion of prediction. Channel capacity of wavebreak locations immediately prior to the phase transition is a  non-zero value. Since the non-zero value is unknown \textit{a priori}, we use the maximum channel capacity (=2.8$\times 10^{-3}$ nats) out of all the wavebreak locations immediately prior to the phase transition (BCL=219 msec) as a cutoff of channel capacity under which a wavebreak can occur. Figure~\ref{fig:error2d}A shows a predicted region of phase transition (white) and wavebreak locations (red) in an error space and a physical space. The error space shows that the predicted region (white) has higher channel capacity than most of the wavebreak locations (Figure~\ref{fig:error2d}A, left two columns). The physical space shows that channel capacity sharply demarcates a circular border across which the wavebreaks occur (Figure~\ref{fig:error2d}A, right column). Of note, the predicted region (white) is located upstream of most of the wavebreaks (red) with respect to the stimulus site. This indicates that channel capacity can define a geometrical border which serves as a frontline of action. Since wavebreaks occur as soon as the wave travels across the border, having the predicted region upstream of wavebreaks has an important therapeutic implication for mitigating the onset of phase transition by modifying the components at the border.

Likewise, we use mutual information as another criterion to predict the region of the phase transition. We use the maximum mutual information (=2.5$\times 10^{-3}$ nats) out of all the wavebreak locations immediately prior to the phase transition (BCL=219 msec) as a cutoff of mutual information under which a wavebreak can occur. As expected from the 1-D and 2-D dynamics of mutual information, the result is almost identical to that of channel capacity in both the error space and the physical space (Figure~\ref{fig:error2d}B).

We also use transfer entropy as an additional criterion of prediction. Since transfer entropy is known to peak prior to a phase transition \citep{barnett2013information}, we used the minimum transfer entropy (=7.3$\times 10^{-4}$ nats) out of all the wavebreak locations immediately prior to the phase transition (BCL=219 msec) as a cutoff of transfer entropy under which a wavebreak can occur. Figure~\ref{fig:error2d}C shows a predicted region of phase transition (white) and wavebreak locations (red) in an error space and a physical space. The error space shows that the predicted region (white) has lower channel capacity than some of the wavebreak locations (Figure~\ref{fig:error2d}C, left two columns). The physical space shows that transfer entropy also demarcates a circular border across which the wavebreaks occur, similar to that of the channel capacity criterion (Figure~\ref{fig:error2d}C, right column). Importantly, the predicted region (white) is immersed in the middle of most of the wavebreaks (red), slightly downstream of the border by the channel capacity criterion. This indicates that, in addition to channel capacity, transfer entropy  can separately define a geometrical border across which phase transitions can occur.

To highlight the difference among those three criteria, Figure~\ref{fig:error2d}D shows the predicted region using either the channel capacity, mutual information, or the transfer entropy criterion. Both the error space (Figure~\ref{fig:error2d}D, left two columns) and the physical space (Figure~\ref{fig:error2d}D, right column) clearly shows that the regions of phase transition predicted by those information metrics match well with the wavebreak locations. However, there is a slight difference between channel capacity/mutual information and transfer entropy in both the error space and the physical space. Importantly, in the physical space, those criteria define different circular borders. This finding indicates that, although both channel capacity/mutual information and transfer entropy can be used to predict phase transitions, they have different underlying information dynamics leading to the transition.

\section*{Discussion}

\subsection*{Summary of main findings}

Using information theory as a guiding principle to describe cardiac dynamics, we successfully simulate an order-disorder phase transition in a cardiac system with an interval between regular stimuli as a control parameter (Figure~\ref{fig:2d_lattice}). Our main findings are summarized as follows:

\begin{enumerate}

\item Channel capacity, mutual information, and transfer entropy can predict geometrical borders beyond which an order-disorder phase transition from a regular heart rhythm to fibrillation occurs in a cardiac system. 

\item Channel capacity and mutual information progressively decline and reach zero at the border of phase transition. This indicates that those information-theoretic metrics can serve as order parameters to describe the macroscopic behavior of the system. Importantly, mutual information can be used as an order parameter in systems where computation of channel capacity is not straightforward because the channel cannot be modeled as a binary asymmetric channel.

\item Mutual information is lower than, but is consistently close to channel capacity, which indicates that the channel is operating close to the maximum capacity, and the cardiac coding sequences are nearly optimized when it is modeled as a binary asymmetric channel.
  
\item Transfer entropy peaks prior to the phase transition at a wide range of BCL. This finding suggests that there is a complex and dynamic information transaction between the stimulus site and individual cardiac components.
  
\item A period-doubling bifurcation precedes the phase transition to fibrillation, and the BCL at bifurcation is spatially heterogeneous due to the node behavior (Figure~\ref{fig:1d_cable},\ref{fig:2d_profile}). This finding confirms that even a simple, anatomically homogeneous model can be functionally heterogeneous in a cardiac system due to intrinsic dynamic properties such as restitution, wavefront curvature and conduction velocity. This finding also demonstrates that functional heterogeneity alone is a sufficient substrate for regional heterogeneity of period-doubling bifurcation, as opposed to anatomical \citep{engelman2010structural} or cellular heterogeneity \citep{diaz2004sarcoplasmic}. In addition, the spatial heterogeneity of APD alternans may account for its lack of sensitivity and specificity as a marker of sudden cardiac death \citep{gold2008role} due to the limitation of accurately quantifying APD alternans using clinically available systems. 
  
\item A period-doubling bifurcation, APD alternans, or even a refractory (2:1) response does not significantly impact the cardiac macrostate. Shannon entropy shows a small and steady decline across the bifurcation as BCL goes down (Figure~\ref{fig:restitution},\ref{fig:dynamics1d},\ref{fig:2d_profile},\ref{fig:bifurcation}). This finding suggests that a period-doubling bifurcation in cardiac systems, although it denotes a qualitative change in the behavior of the system as a function of BCL, does not change macroscopic behaviors. This further provides a support that a period-doubling bifurcation and APD alternans alone do not contribute to an order-disorder phase transition from a regular heart rhythm to fibrillation in a cardiac system.
 
\end{enumerate}

\subsection*{Information-theoretic analysis to locate order-disorder phase transition}

Previous studies report that mutual information between all-to-all pairs of components in the system peaks precisely at the phase transition in many complex systems, including the Ising model \citep{matsuda1996mutual,gu2007universal}, a swarm model \citep{vicsek1995novel,wicks2007mutual}, random Boolean networks \citep{ribeiro2008mutual}, and financial markets \citep{harre2009phase}. When the system is highly ordered, little uncertainty about the state of individual components makes mutual information small. In contrast, when the system is highly disordered, mutual information is also small because the components behave almost independently. The mutual information is maximum at the phase transition where susceptibility peaks\citep{barnett2013information}. In contrast, in this work mutual information and channel capacity do not peak at the order-disorder phase transition. Instead, they change from non-zero to zero at the transition from the ordered side to the disordered side. This is because those metrics in our approach are essentially evaluating how well two components are communicating with each other. This novel approach is characterized by three advantages. First, it allows channel capacity and mutual information as order parameters to quantify the macroscopic behaviors of the system. The phase transition occurs at the weakest link of communication, or miscommunication. Second, those information-theoretic metrics help predict order-disorder phase transitions, since they become zero at the transition. This is particularly important since mutual information traditionally does not help predict phase transitions since it peaks precisely at transition. Third, it is computationally efficient even in a system with large numbers of components.

The role of transfer entropy in predicting order-disorder phase transitions is less characterized than that of mutual information. In random Boolean networks, a phase transition is characterized by the shifting balance of local information storage over transfer\citep{lizier2008information}. In the Ising model, a collective multivariate transfer entropy, called global transfer entropy, peaks on the disordered side of a transition\citep{barnett2013information}. Global transfer entropy indicates a balance between integration and segregation of complex networks of dynamical processes\citep{barrett2010multivariate,seth2005causal,tononi1994measure}. The mechanism of these results remains unclear, but it is relevant to many natural and social systems where disorder is associated with healthy features and order with pathological dynamics, such as synchronization in epileptic seizures and herding behaviors in financial market crashes\citep{Bossomaier2016transferentropy}. In contrast, in the heart, order is associated with healthy dynamics and disorder is associated with pathological, often lethal dynamics. In our data, we find that the transfer entropy peaks on the ordered side of a transition. In particular, transfer entropy peaks beyond the period-doubling bifurcation are found to help identify the locations of phase transition initiation. It is possible that transfer entropy peak between the period-doubling bifurcation and the phase transition at wavebreaks uncovers complex interactions between the driving component and the local components at wavebreaks near criticality.

\subsection*{Clinical implications}
Implantable cardioverter-defibrillators (ICD) are the standard of care for primary prevention of sudden cardiac death in high-risk patients who are yet to experience fatal events \citep{epstein20132012}. However, only a minority of ICD recipients experience appropriate firings based on the current criteria for primary prevention ICD, relying on impaired cardiac function assessed by functional class and left ventricular ejection fraction \citep{bardy2005amiodarone}. In addition, the cost and the risks of complications \citep{ranasinghe2016long} and inappropriate firings \citep{van2011inappropriate} do not warrant indiscriminate application of ICD therapy. Our approach provides an alternative strategy by identifying and targeting the cardiac components with interventional catheter ablation therapies to prevent sudden death resulting from ventricular fibrillation \citep{knecht2009long,cheniti2017vf,krummen2015modifying}. Our approach is potentially applicable to other clinical conditions such as epileptic seizures\citep{viventi2011flexible} in which spiral waves play a major role. Application of this approach could also have relevance to a wide range of systems to mitigate or prevent the imminent order-disorder phase transition by identifying and eliminating the component responsible for the initiation of the transition.

\subsection*{Limitations}
We recognize several limitations associated with the numerical method we implemented. First, we use the Fenton-Karma model \citep{fenton1998vortex}, which is a relatively simple cardiac model, with a homogeneous and isotropic 2-D lattice. It is possible that a more biophysically detailed model of the heart with anatomical heterogeneity, anisotropy and a more realistic geometry could make our approach more difficult to analyze. For example, the functional heterogeneity that we observe may further be accentuated in a more realistic model, making the analysis more vulnerable to noise. However, the simplicity of the cardiac model is an advantage that allows the results from this model to be widely applicable to other complex systems to gain general insights as to how information-theoretic metrics help locate phase transitions. 

\subsection*{Conclusions}
We developed an information-theoretic approach to predict the locations of an order-disorder phase transition from a regular heart rhythm to fibrillation in a cardiac system. Our computationally efficient approach is applicable not only to the cardiac system but also to a wide range of systems in distinct physical, chemical and biological systems.

\matmethods{}
\showmatmethods 
We performed the simulation and the data analysis using Matlab R2017a (Mathworks, Inc.).

\subsection*{Model of cardiac system}
We used a simplified mathematical ionic model of the cardiac action potential described by Fenton and Karma~\citep{fenton1998vortex}. We chose this model because it accurately reproduces the critical properties of the cardiac action potential to test our hypothesis, such as restitution properties, APD alternans, conduction block, and spiral wave initiation ~\citep{fenton2002multiple}. The model consists of three variables: the transmembrane potential $V$, a fast ionic gate $u$, and a slow ionic gate $w$.

\begin{equation}
  \frac{\partial V}{\partial t} = \nabla\cdot (D\nabla V) - \frac{I_{fi}+I_{so}+I_{si}+I_{ex}}{C_m}\\
  \label{eq:FK01}
\end{equation}

Here $C_m$ is the membrane capacitance (= 1 $\mu F/cm^2$), and $D$ is the diffusion tensor, which is a diagonal matrix whose diagonal and off-diagonal elements are 0.001 cm$^2$/msec and 0 cm$^2$/msec, respectively, to represent a 2-D isotropic system\citep{fenton2002multiple}. The current $I_{fi}$ is a fast inward inactivation current used to depolarize the membrane when an excitation above threshold is induced. The current $I_{so}$ is a slow, time-independent rectifying outward current used to repolarize the membrane back to the resting potential. The current $I_{si}$ is a slow inward inactivation current used to balance $I_{so}$ and to produce the observed plateau in the action potential. $I_{ex}$ is the external current~\citep{pertsov1993spiral}. The two gate variables of the model follow first order equations in time.
\begin{align}
  \frac{\partial u}{\partial t} &= \frac{(1-p)(1-u)}{\tau^{-}_{u}}-\frac{pu}{\tau^{+}_{u}}\\
  \frac{\partial w}{\partial t} &= \frac{(1-p)(1-w)}{\tau^{-}_{w}}-\frac{pw}{\tau^{+}_{w}}
\label{eq:FK02}
\end{align}
where
\begin{equation}
\tau^{-}_{u}=(1-q)\tau^{-}_{u1}+q\tau^-_{u2}
  \label{eq:FK03}
\end{equation}
and
\begin{equation}
p =\begin{cases}
1 & \text{ if } V\geq V_c \\ 
0 & \text{ if } V< V_c 
\end{cases}
  \label{eq:FK04}
\end{equation}
\begin{equation}
q =\begin{cases}
1 & \text{ if } V\geq V_u \\ 
0 & \text{ if } V< V_u 
\end{cases}
  \label{eq:FK05}
\end{equation}
The transmembrane potential $V$ and the two gate variables $u$ and $w$ vary from 0 to 1. The three currents are given by the following.
\begin{align}
  I_{fi} &= -\frac{up}{\tau_d}(V-V_c)(1-V)\\
  I_{so} &= \frac{V}{\tau_0}(1-p)+\frac{p}{\tau_r}\\
  I_{si} &= -\frac{w}{2\tau_{si}}(1+\tanh(k(V-V^{si}_c))) 
\label{eq:FK06}
\end{align}
We chose the following model parameters to produce action potential dynamics of interest for the study, including APD alternans, conduction block and wavebreak generation: $\tau^{+}_u$=3.33; $\tau^{-}_{u1}$=15.6; $\tau^{-}_{u2}$=5; $\tau^{+}_{w}$=350; $\tau^{-}_{w}$=80; $\tau_{d}$=0.407; $\tau_{0}$=9; $\tau_{r}$=34; $\tau_{si}$=26.5; k=15; $V^{si}_{c}$=0.45; $V_{c}$=0.15; and $V_{u}$=0.04~\citep{fenton2002multiple}. 

We solved the model equations with an explicit Euler integration at $\Delta t$=0.1 msec. For 1-D and 2-D simulations, we used a finite difference method for spatial derivatives ($\Delta x$=0.025 cm) assuming Neumann boundary conditions. This set of parameters satisfies von Neumann stability requirement for 1-D and 2-D finite difference schemes.

\begin{equation}
\frac{D\Delta t}{(\Delta x)^2}\leq \frac{1}{2d}
\label{eq:stability}
\end{equation}

where $d=1$ for 1-D and $d=2$ for 2-D simulations.

The interval between regular stimuli with an external current, or basic cycle length (BCL), ranged between 200 msec and 300 msec. We applied 220 stimuli for 1-D and 60 stimuli for 2-D simulations at each BCL to reach a steady state prior to progressively reducing the BCL by a decrement of 1 msec to achieve the minimum effective refractory period. 

\subsection*{Cardiac simulation}
For 1-D simulations we stimulated the origin ($x=0$) of the system of a 1-D cable of length 40 cm and acquired the time series of 60 beats in each component at each BCL. For 2-D simulations we stimulated the top left component of the system of a 2-D lattice of 25 cm $\times$ 25 cm and acquired the time series of 60 beats in each component at each BCL. The time series were downsampled to achieve the final sampling frequency of 1 kHz ($\Delta t$=1 msec) to reflect realistic measurements in human clinical electrophysiology studies~\citep{fogoros2012electrophysiologic}.

\subsection*{Restitution properties}
We calculated APD, CV and diastolic interval (DI) at each BCL several points away from the stimulus site in the 1-D cable to avoid stimulus artifacts and the boundary effects~\citep{fenton2002multiple}. APD was measured at 90$\%$ repolarization ($APD_{90}$) at a steady state at each component. The $i$th diastolic interval ($DI_{i}$) is defined as the difference between BCL and the $i$th $APD$ ($APD_i$) (Figure~\ref{fig:conceptual}A):
\begin{equation}
DI_{i} = BCL - APD_{i}
  \label{eq:DI}
\end{equation}
We used local regression with weighted linear least squares and a first-degree polynomial model to determine CV at each BCL, because it depends sensitively on the time measurements~\citep{fenton2002multiple}. 

\subsection*{APD alternans, curvature, and wavebreak}
For both 1-D and 2-D simulations, the period-doubling bifurcation was defined as the longest BCL that generates APD alternans at a steady state. APD alternans was defined to be present when APD alternates between the longer and the shorter APD in response to regular stimuli, and the difference between the longer and the shorter APD is greater than 5 msec. For 2-D simulations, curvature of the wavefront was defined as the reciprocal of the radius of its osculating circle at each component and was described in arbitrary units. When the stimuli generate wavebreaks that initiate spiral waves, we identified the coordinates of wavebreaks in the 2-D lattice. The wavebreak was defined as the point of maximum curvature intersecting the wavefront and the waveback that initiates the first spiral wave in each time series. There are typically 2-6 wavebreaks per time series.

\subsection*{Information-theoretic metrics}
The time series of cardiac excitation in each component was encoded as 1 when excited (during $APD_{90}$) and 0 when resting\citep{ashikaga2015modelling} (Figure~\ref{fig:conceptual}A). The conduction delay to travel from the stimulus site to each component was subtracted from the time series to allow comparison of the same traveling wave at each component. We treated each component as a time-series process $X$, where at any observation time $t$ the process $X$ is either excited ($X=1$) or resting ($X=0$). We consider the stimulus site as an information sender/encoder with an input $X$, and any other component in the system as a receiver/decoder with an output $Y$ (Figure~\ref{fig:conceptual}B). We consider the intervening components between the sender/encoder and the receiver/decoder as channels. 

The Shannon entropy $H$ of each time-series process $X$ is given by
\begin{eqnarray}
H(X)&=&-\sum_{x}p(x)\log_{2}p(x)\\
&=&-p(X=0)\log_{2}p(X=0)-p(X=1)\log_{2}p(X=1)
\label{eq:entropy01}
\end{eqnarray}
where $p(x)$ denotes the probability density function of the time series generated by $X$. This quantifies the average uncertainty of whether a single component is excited ($x=1$) or resting ($x=0$) over the time history \cite{2006:CoverEIT}. 

The channel capacity $C$ between the input $X$ and the output $Y$ is given by
\begin{equation}
C=\max_{p(x)}I(X;Y)
  \label{eq:Cap2}
\end{equation}
where $I(X;Y)$ is the mutual information that quantifies the information content shared between the input $X$ and the output $Y$.
\begin{equation}
I(X;Y)=\sum_{x,y}p(x,y)\log_{2}\frac{p(x,y)}{p(x)p(y)}
  \label{eq:mi}
\end{equation}
where $p(x,y)$ denotes the joint probability density function of $X$ and $Y$. We consider these channels to be a binary asymmetric channel, which is the most general form of binary discrete memoryless channel (Figure~\ref{fig:conceptual}C). The channel has a probability $\varepsilon_0$ that an input 0 will be flipped into a 1 (type 0 error) and a probability $\varepsilon_1$ for a flip from 1 to 0 (type 1 error). 
\begin{align}
p(Y=0|X=0)&=1-\varepsilon_0\\
p(Y=1|X=0)&=\varepsilon_0\\
p(Y=1|X=1)&=1-\varepsilon_1\\
p(Y=0|X=1)&=\varepsilon_1
\label{eq:errors}
\end{align}

If $p(X=0)=x$ and $p(X=1)=1-x$, 
\begin{align}
p(Y=0)&=(1-\varepsilon_0)p(X=0)+\varepsilon_1p(X=1)\\
&=(1-\varepsilon_0-\varepsilon_1)x+\varepsilon_1
\label{eq:pY}
\end{align}
The conditional entropy of $Y$ given $X$ is 
\begin{equation}
H(Y|X)=xh(\varepsilon_0)+(1-x)h(\varepsilon_1)
\end{equation}
where $h$ is the binary entropy function defined as
\begin{equation}
h(p)=-p\log_2 p-(1-p)\log_2 (1-p)
  \label{eq:bient}
\end{equation}

The mutual information as a function of $x$ is 
\begin{align}
I(x)&=I(X;Y)=H(Y)-H(Y|X)\\
&=h[(1-\varepsilon_0-\varepsilon_1)x+\varepsilon_1]-xh(\varepsilon_0)-(1-x)h(\varepsilon_1)\\
&=h[(1-\varepsilon_0-\varepsilon_1)x+\varepsilon_1]-x[h(\varepsilon_0)-h(\varepsilon_1)]-h(\varepsilon_1)
\end{align}
The derivative of $I(x)$ is
\begin{equation}
I'(x)=(1-\varepsilon_0-\varepsilon_1)\log_2\left[\frac{1}{x(1-\varepsilon_0-\varepsilon_1)+\varepsilon_1}-1\right]-\left[h(\varepsilon_0)-h(\varepsilon_1)\right] 
\end{equation}
because
\begin{equation}
h'(x)=\log_2\left(\frac{1}{x}-1\right)
\end{equation}
Solving $I(x)=0$ gives
\begin{equation}
\frac{1}{x(1-\varepsilon_0-\varepsilon_1)+\varepsilon_1}-1=z
\end{equation}
where
\begin{equation}
z=2^{\frac{h(\varepsilon_0)-h(\varepsilon_1)}{1-\varepsilon_0-\varepsilon_1}}
\end{equation}
and thus
\begin{align}
x&=\frac{1}{1-\varepsilon_0-\varepsilon_1}\left(\frac{1}{z+1}-\varepsilon_1  \right)\\
&=\frac{1-\varepsilon_1(1+z)}{(1-\varepsilon_0-\varepsilon_1)(1+z)}
\end{align}
Therefore, the channel capacity of the binary asymmetric channel is
\begin{align}
C&=h\left[\frac{1-\varepsilon_1(1+z)}{1+z}+\varepsilon_1 \right]-\frac{1-\varepsilon_1(1+z)}{(1-\varepsilon_0-\varepsilon_1)(1+z)}\left[h(\varepsilon_0)-h(\varepsilon_1)\right]-h(\varepsilon_1)\\
&=h\left(\frac{1}{1+z}\right)-\frac{\log_2z}{1+z}+\varepsilon_1\log_2z-h(\varepsilon_1)
\end{align}
because
\begin{equation}
h\left(\frac{1}{1+z}\right)=\log_2(1+z)-\frac{z\log_2z}{z+1}
\end{equation}
$C$ can be rewritten as
\begin{align}
C&=\log_2(1+z)-\log_2z+\varepsilon_1\log_2z-h(\varepsilon_1)\\
&=\log_2(1+z)-\frac{1-\varepsilon_1}{1-\varepsilon_0-\varepsilon_1}\left[h(\varepsilon_0)-h(\varepsilon_1)\right]-h(\varepsilon_1)\\
&=\log_2(1+z)-\frac{1-\varepsilon_1}{1-\varepsilon_0-\varepsilon_1}h(\varepsilon_0)+\frac{\varepsilon_0}{1-\varepsilon_0-\varepsilon_1}h(\varepsilon_1)
\end{align}

The input distribution $p(x)$ that maximizes the mutual information and thus achieves the channel capacity is given by\citep{moser2009error}
\begin{equation}
p(X=0)=1-p(X=1)=\frac{1-\varepsilon_1(1+z)}{(1-\varepsilon_0-\varepsilon_1)(1+z)}
  \label{eq:inputD}
\end{equation}

The transfer entropy~\citep{schreiber2000measuring} is a non-parametric statistic measuring the directed reduction in uncertainty in one time-series given another, generally interpreted as information transfer. The transfer entropy from the input $X$ to the output $Y$ is the amount of uncertainty reduced in future values of $Y$ by knowing the past values of $X$, given past values of $Y$ \citep{ashikaga2017hidden}.
\begin{align}
T_{X \rightarrow Y}
  &= \sum p(y_{t+1},y^l_t,x^k_t) \log_2 \frac{p(y_{t+1}|y^l_t,x^k_t)}{p(y_{t+1}|y_t^l)} \\
  &= H(y_{t+1}|y^l_t) - H(y_{t+1}|y^l_t,x^k_t)
  ~,
  \label{eq:Txy}
\end{align}
where $k$ and $l$ denote the length of time series in the processes $X$ and $Y$, respectively:
\begin{align}
  x^k_t &= (x_t,x_{t-1},...,x_{t-k+1}) \\
  y^l_t &= (y_t,y_{t-1},...,y_{t-l+1})
  ~.
  \label{eq:xkyl}
\end{align}
We define $k$ and $l$ such that $x^k_t$ and $y^l_t$ contain the entire time-series ($k=l$). We used the discrete transfer entropy calculator of the Java Information Dynamics Toolkit (JIDT) to calculate transfer entropy~\citep{lizier2014jidt}.

 We adopt the standard convention of $0 \cdot \log_2{0} = 0$. 

\acknow{This work was supported by the Fondation Leducq Transatlantic Network of Excellence.}

\showacknow 


\bibliography{alternans_ref}

\begin{thebibliography}{10}

\bibitem{buhl2006disorder}
Buhl J, et~al. (2006) From disorder to order in marching locusts.
\newblock {\em Science} 312(5778):1402--1406.

\bibitem{miramontes1995order}
Miramontes O (1995) Order-disorder transitions in the behavior of ant
  societies.
\newblock {\em Complexity} 1(3):56--60.

\bibitem{may2008complex}
May RM, Levin SA, Sugihara G (2008) Complex systems: Ecology for bankers.
\newblock {\em Nature} 451(7181):893--895.

\bibitem{lenton2008tipping}
Lenton TM, et~al. (2008) Tipping elements in the earth's climate system.
\newblock {\em Proceedings of the national Academy of Sciences}
  105(6):1786--1793.

\bibitem{scheffer2001catastrophic}
Scheffer M, Carpenter S, Foley JA, Folke C, Walker B (2001) Catastrophic shifts
  in ecosystems.
\newblock {\em Nature} 413(6856):591--596.

\bibitem{dai2012generic}
Dai L, Vorselen D, Korolev KS, Gore J (2012) Generic indicators for loss of
  resilience before a tipping point leading to population collapse.
\newblock {\em Science} 336(6085):1175--1177.

\bibitem{dai2013slower}
Dai L, Korolev KS, Gore J (2013) Slower recovery in space before collapse of
  connected populations.
\newblock {\em Nature} 496(7445):355--358.

\bibitem{kramer2012human}
Kramer MA, et~al. (2012) Human seizures self-terminate across spatial scales
  via a critical transition.
\newblock {\em Proceedings of the National Academy of Sciences}
  109(51):21116--21121.

\bibitem{Christodoulidi2014flocking}
Christodoulidi H, van~der Weele K, Antonopoulos CG, Bountis T (2014) {\em Phase
  transitions in models of bird flocking In Chaos, Information Processing and
  Paradoxical Games: The Legacy of John S Nicolis}, eds.{} Nicolis G, Basios V.
\newblock (World Scientific Publishing Company), pp. 383--98.

\bibitem{mora2011biological}
Mora T, Bialek W (2011) Are biological systems poised at criticality?
\newblock {\em Journal of Statistical Physics} 144(2):268--302.

\bibitem{hesse2014self}
Hesse J, Gross T (2014) Self-organized criticality as a fundamental property of
  neural systems.
\newblock {\em Frontiers in systems neuroscience} 8.

\bibitem{hidalgo2014information}
Hidalgo J, et~al. (2014) Information-based fitness and the emergence of
  criticality in living systems.
\newblock {\em Proceedings of the National Academy of Sciences}
  111(28):10095--10100.

\bibitem{scheffer2009early}
Scheffer M, et~al. (2009) Early-warning signals for critical transitions.
\newblock {\em Nature} 461(7260):53--59.

\bibitem{hayashi2015spectrum}
Hayashi M, Shimizu W, Albert CM (2015) The spectrum of epidemiology underlying
  sudden cardiac death.
\newblock {\em Circulation research} 116(12):1887--1906.

\bibitem{haissaguerre1998spontaneous}
Haissaguerre M, et~al. (1998) Spontaneous initiation of atrial fibrillation by
  ectopic beats originating in the pulmonary veins.
\newblock {\em New England Journal of Medicine} 339(10):659--666.

\bibitem{haissaguerre2016ventricular}
Haissaguerre M, Vigmond E, Stuyvers B, Hocini M, Bernus O (2016) Ventricular
  arrhythmias and the his-purkinje system.
\newblock {\em Nature reviews. Cardiology} 13(3):155.

\bibitem{verma2015approaches}
Verma A, et~al. (2015) Approaches to catheter ablation for persistent atrial
  fibrillation.
\newblock {\em New England Journal of Medicine} 372(19):1812--1822.

\bibitem{knecht2009long}
Knecht S, et~al. (2009) Long-term follow-up of idiopathic ventricular
  fibrillation ablation.
\newblock {\em Journal of the American College of Cardiology} 54(6):522--528.

\bibitem{cheniti2017vf}
Cheniti G, et~al. (2017) Is vf an ablatable rhythm?
\newblock {\em Current treatment options in cardiovascular medicine} 19(2):14.

\bibitem{krummen2015modifying}
Krummen DE, et~al. (2015) Modifying ventricular fibrillation by targeted rotor
  substrate ablation: proof-of-concept from experimental studies to clinical
  vf.
\newblock {\em Journal of cardiovascular electrophysiology} 26(10):1117--1126.

\bibitem{quax2013diminishing}
Quax R, Apolloni A, Sloot PM (2013) The diminishing role of hubs in dynamical
  processes on complex networks.
\newblock {\em Journal of The Royal Society Interface} 10(88):20130568.

\bibitem{1948:shannon01}
Shannon CE (1948) A mathematical theory of communication.
\newblock {\em Bell Syst Tech J} 27(379-423):623--656.

\bibitem{schreiber2000measuring}
Schreiber T (2000) Measuring information transfer.
\newblock {\em Physical review letters} 85(2):461.

\bibitem{matsuda1996mutual}
Matsuda H, Kudo K, Nakamura R, Yamakawa O, Murata T (1996) Mutual information
  of ising systems.
\newblock {\em International Journal of Theoretical Physics} 35(4):839--845.

\bibitem{gu2007universal}
Gu SJ, Sun CP, Lin HQ (2007) Universal role of correlation entropy in critical
  phenomena.
\newblock {\em Journal of Physics A: Mathematical and Theoretical}
  41(2):025002.

\bibitem{vicsek1995novel}
Vicsek T, Czir{\'o}k A, Ben-Jacob E, Cohen I, Shochet O (1995) Novel type of
  phase transition in a system of self-driven particles.
\newblock {\em Physical review letters} 75(6):1226.

\bibitem{wicks2007mutual}
Wicks RT, Chapman SC, Dendy R (2007) Mutual information as a tool for
  identifying phase transitions in dynamical complex systems with limited data.
\newblock {\em Physical Review E} 75(5):051125.

\bibitem{ribeiro2008mutual}
Ribeiro AS, Kauffman SA, Lloyd-Price J, Samuelsson B, Socolar JE (2008) Mutual
  information in random boolean models of regulatory networks.
\newblock {\em Physical Review E} 77(1):011901.

\bibitem{harre2009phase}
Harr{\'e} M, Bossomaier T (2009) Phase-transition--like behaviour of
  information measures in financial markets.
\newblock {\em EPL (Europhysics Letters)} 87(1):18009.

\bibitem{barnett2013information}
Barnett L, Lizier JT, Harr{\'e} M, Seth AK, Bossomaier T (2013) Information
  flow in a kinetic ising model peaks in the disordered phase.
\newblock {\em Physical review letters} 111(17):177203.

\bibitem{weiss2005dynamics}
Weiss JN, et~al. (2005) The dynamics of cardiac fibrillation.
\newblock {\em Circulation} 112(8):1232--1240.

\bibitem{ashikaga2015modelling}
Ashikaga H, et~al. (2015) Modelling the heart as a communication system.
\newblock {\em Journal of The Royal Society Interface} 12(105):20141201.

\bibitem{guevara1984electrical}
Guevara M, Ward G, Shrier A, Glass L (1984) Electrical alternans and period
  doubling bifurcations.
\newblock {\em IEEE Comp Cardiol} 562:167--170.

\bibitem{quail2015predicting}
Quail T, Shrier A, Glass L (2015) Predicting the onset of period-doubling
  bifurcations in noisy cardiac systems.
\newblock {\em Proceedings of the National Academy of Sciences}
  112(30):9358--9363.

\bibitem{restrepo2008calsequestrin}
Restrepo JG, Weiss JN, Karma A (2008) Calsequestrin-mediated mechanism for
  cellular calcium transient alternans.
\newblock {\em Biophysical journal} 95(8):3767--3789.

\bibitem{restrepo2009spatiotemporal}
Restrepo JG, Karma A (2009) Spatiotemporal intracellular calcium dynamics
  during cardiac alternans.
\newblock {\em Chaos: An Interdisciplinary Journal of Nonlinear Science}
  19(3):037115.

\bibitem{rovetti2010spark}
Rovetti R, Cui X, Garfinkel A, Weiss JN, Qu Z (2010) Spark-induced sparks as a
  mechanism of intracellular calcium alternans in cardiac myocytes.
\newblock {\em Circulation research} 106(10):1582--1591.

\bibitem{alvarez2015calcium}
Alvarez-Lacalle E, Echebarria B, Spalding J, Shiferaw Y (2015) Calcium
  alternans is due to an order-disorder phase transition in cardiac cells.
\newblock {\em Physical review letters} 114(10):108101.

\bibitem{rosenbaum1994electrical}
Rosenbaum DS, et~al. (1994) Electrical alternans and vulnerability to
  ventricular arrhythmias.
\newblock {\em New England Journal of Medicine} 330(4):235--241.

\bibitem{gold2008role}
Gold MR, et~al. (2008) Role of microvolt t-wave alternans in assessment of
  arrhythmia vulnerability among patients with heart failure and systolic
  dysfunction.
\newblock {\em Circulation} 118(20):2022--2028.

\bibitem{goldberger2016sublimation}
Goldberger AL, Henriques TS, Mariani S (2016) Sublimation-like behavior of
  cardiac dynamics in heart failure: A malignant phase transition?
\newblock {\em Complexity} 21(S2):24--32.

\bibitem{2014zipes:aa}
Gray RA (2014) {\em Theory of rotors and arrhythmias. In Cardiac
  Electrophysiology: From Cell to Bedside}, eds.{} Zipes DP, Jalife J.
\newblock (Saunders), 6 edition, pp. 191--223.

\bibitem{cabo1994wave}
Cabo C, et~al. (1994) Wave-front curvature as a cause of slow conduction and
  block in isolated cardiac muscle.
\newblock {\em Circulation research} 75(6):1014--1028.

\bibitem{fenton2002multiple}
Fenton FH, Cherry EM, Hastings HM, Evans SJ (2002) Multiple mechanisms of
  spiral wave breakup in a model of cardiac electrical activity.
\newblock {\em Chaos: An Interdisciplinary Journal of Nonlinear Science}
  12(3):852--892.

\bibitem{engelman2010structural}
Engelman ZJ, Trew ML, Smaill BH (2010) Structural heterogeneity alone is a
  sufficient substrate for dynamic instability and altered restitution.
\newblock {\em Circulation: Arrhythmia and Electrophysiology} pp. CIRCEP--109.

\bibitem{diaz2004sarcoplasmic}
D{\'\i}az ME, O'Neill SC, Eisner DA (2004) Sarcoplasmic reticulum calcium
  content fluctuation is the key to cardiac alternans.
\newblock {\em Circulation research} 94(5):650--656.

\bibitem{lizier2008information}
Lizier JT, Prokopenko M, Zomaya AY (2008) The information dynamics of phase
  transitions in random boolean networks. in {\em ALIFE}.
\newblock pp. 374--381.

\bibitem{barrett2010multivariate}
Barrett AB, Barnett L, Seth AK (2010) Multivariate granger causality and
  generalized variance.
\newblock {\em Physical Review E} 81(4):041907.

\bibitem{seth2005causal}
Seth AK (2005) Causal connectivity of evolved neural networks during behavior.
\newblock {\em Network: Computation in Neural Systems} 16(1):35--54.

\bibitem{tononi1994measure}
Tononi G, Sporns O, Edelman GM (1994) A measure for brain complexity: relating
  functional segregation and integration in the nervous system.
\newblock {\em Proceedings of the National Academy of Sciences}
  91(11):5033--5037.

\bibitem{Bossomaier2016transferentropy}
Bossomaier T, Barnett L, Harr{\'e} M, Lizier J, eds. (2016) {\em An
  Introduction to Transfer Entropy}.
\newblock (Springer), pp. 65--96.

\bibitem{epstein20132012}
Epstein AE, et~al. (2013) 2012 accf/aha/hrs focused update incorporated into
  the accf/aha/hrs 2008 guidelines for device-based therapy of cardiac rhythm
  abnormalities.
\newblock {\em Circulation} 127(3):e283--e352.

\bibitem{bardy2005amiodarone}
Bardy GH, et~al. (2005) Amiodarone or an implantable
  cardioverter--defibrillator for congestive heart failure.
\newblock {\em New England Journal of Medicine} 352(3):225--237.

\bibitem{ranasinghe2016long}
Ranasinghe I, et~al. (2016) Long-term risk for device-related complications and
  reoperations after implantable cardioverter-defibrillator implantationan
  observational cohort studylong-term nonfatal outcomes after icd implantation.
\newblock {\em Annals of internal medicine} 165(1):20--29.

\bibitem{van2011inappropriate}
van Rees JB, et~al. (2011) Inappropriate implantable cardioverter-defibrillator
  shocks: incidence, predictors, and impact on mortality.
\newblock {\em Journal of the American College of Cardiology} 57(5):556--562.

\bibitem{viventi2011flexible}
Viventi J, et~al. (2011) Flexible, foldable, actively multiplexed, high-density
  electrode array for mapping brain activity in vivo.
\newblock {\em Nature neuroscience} 14(12):1599--1605.

\bibitem{fenton1998vortex}
Fenton F, Karma A (1998) Vortex dynamics in three-dimensional continuous
  myocardium with fiber rotation: filament instability and fibrillation.
\newblock {\em Chaos: An Interdisciplinary Journal of Nonlinear Science}
  8(1):20--47.

\bibitem{pertsov1993spiral}
Pertsov AM, Davidenko JM, Salomonsz R, Baxter WT, Jalife J (1993) Spiral waves
  of excitation underlie reentrant activity in isolated cardiac muscle.
\newblock {\em Circulation research} 72(3):631--650.

\bibitem{fogoros2012electrophysiologic}
Fogoros RN (2012) {\em Electrophysiologic testing}.
\newblock (John Wiley \&amp; Sons).

\bibitem{2006:CoverEIT}
Cover TM, Thomas JA (2006) {\em Elements of Information Theory}.
\newblock (Wiley-Interscience), 2 edition.

\bibitem{moser2009error}
Moser SM (2009) Error probability analysis of binary asymmetric channels.
\newblock {\em Dept. El. \&amp; Comp. Eng., Nat. Chiao Tung Univ}.

\bibitem{ashikaga2017hidden}
Ashikaga H, James RG (2017) Hidden structures of information transport
  underlying spiral wave dynamics.
\newblock {\em Chaos: An Interdisciplinary Journal of Nonlinear Science}
  27(1):013106.

\bibitem{lizier2014jidt}
Lizier JT (2014) Jidt: An information-theoretic toolkit for studying the
  dynamics of complex systems.
\newblock {\em Frontiers in Robotics and AI} 1:11.

\end{thebibliography}

\end{document}